\PassOptionsToPackage{unicode}{hyperref}
\PassOptionsToPackage{hyphens}{url}
\PassOptionsToPackage{dvipsnames,svgnames,x11names}{xcolor}
\documentclass[
  table]{article}
\usepackage{amsmath,amssymb}
\usepackage{iftex}
\ifPDFTeX
  \usepackage[T1]{fontenc}
  \usepackage[utf8]{inputenc}
  \usepackage{textcomp} 
\else 
  \usepackage{unicode-math} 
  \defaultfontfeatures{Scale=MatchLowercase}
  \defaultfontfeatures[\rmfamily]{Ligatures=TeX,Scale=1}
\fi
\usepackage{lmodern}
\ifPDFTeX\else
\fi
\IfFileExists{upquote.sty}{\usepackage{upquote}}{}
\IfFileExists{microtype.sty}{
  \usepackage[]{microtype}
  \UseMicrotypeSet[protrusion]{basicmath} 
}{}
\makeatletter
\@ifundefined{KOMAClassName}{
  \IfFileExists{parskip.sty}{%
    \usepackage{parskip}
  }{
    \setlength{\parindent}{0pt}
    \setlength{\parskip}{6pt plus 2pt minus 1pt}}
}{
  \KOMAoptions{parskip=half}}
\makeatother
\usepackage{xcolor}
\usepackage{longtable,booktabs,array}
\usepackage{calc} 
\usepackage{etoolbox}
\makeatletter
\patchcmd\longtable{\par}{\if@noskipsec\mbox{}\fi\par}{}{}
\makeatother
\IfFileExists{footnotehyper.sty}{\usepackage{footnotehyper}}{\usepackage{footnote}}
\makesavenoteenv{longtable}
\usepackage{graphicx}
\makeatletter
\def\maxwidth{\ifdim\Gin@nat@width>\linewidth\linewidth\else\Gin@nat@width\fi}
\def\maxheight{\ifdim\Gin@nat@height>\textheight\textheight\else\Gin@nat@height\fi}
\makeatother
\setkeys{Gin}{width=\maxwidth,height=\maxheight,keepaspectratio}
\makeatletter
\def\fps@figure{htbp}
\makeatother
\providecommand{\tightlist}{%
  \setlength{\itemsep}{0pt}\setlength{\parskip}{0pt}}
\newlength{\cslhangindent}
\newlength{\csllabelwidth}
\newlength{\cslentryspacingunit} 
\newenvironment{CSLReferences}[2] 
 {
  \setlength{\parindent}{0pt}
  \ifodd #1
  \let\oldpar\par
  \def\par{\hangindent=\cslhangindent\oldpar}
  \fi
  \setlength{\parskip}{#2\cslentryspacingunit}
 }%
 {}
\usepackage{calc}

\usepackage{hyperref}
\rowcolors{2}{white}{gray!25}
\def\BibTeX{{\rm B\kern-.05em{\sc i\kern-.025em b}\kern-.08emT\kern-.1667em\lower.7ex\hbox{E}\kern-.125emX}}
\ifLuaTeX
  \usepackage{selnolig}  
\fi
\IfFileExists{bookmark.sty}{\usepackage{bookmark}}{\usepackage{hyperref}}
\IfFileExists{xurl.sty}{\usepackage{xurl}}{} 
\hypersetup{
  pdftitle={DatAasee: A Metadata-Lake as Metadata Catalog for a Virtual Data-Lake},
  pdfauthor={Christian Himpe},
  pdfkeywords={Research Data Management, Metadata Catalog, Metadata
Aggregator, Metadata Lake, Library Science},
  colorlinks=true,
  linkcolor={blue},
  filecolor={Maroon},
  citecolor={Blue},
  urlcolor={Blue},
  pdfcreator={LaTeX via pandoc}}

\title{\texttt{DatAasee}: A Metadata-Lake as Metadata Catalog for a
Virtual Data-Lake}
\author{Christian Himpe\footnote{University and State Library of
  Münster, University of Münster, Krummer Timpen 3, 48143 Münster,
  Germany.
  \href{mailto:christian.himpe@uni-muenster.de}{\nolinkurl{christian.himpe@uni-muenster.de}}
  ,
  \href{https://orcid.org/0000-0003-2194-6754}{orcid:0000-0003-2194-6754}}}
\date{}

\begin{document}
\maketitle
\begin{abstract}
Metadata management for distributed data sources is a long-standing but
ever-growing problem. To counter this challenge in a research-data and
library-oriented setting, this work constructs a data architecture,
derived from the data-lake: the metadata-lake. A proof-of-concept
implementation of this proposed metadata aggregator is presented and
briefly evaluated.
\end{abstract}

\hypertarget{introduction}{%
\section{Introduction}\label{introduction}}

In the shadows of data evolving to big data, metadata became big
metadata. And due to metadata's idiosyncrasy of describing other data,
metadata management for big data warrants its own specialized
discussion. In particular the inert dependency of metadata on its data
needs to be accounted for, and leads to the driving conjecture behind
this work and its object of study: existence, detail and accuracy of
metadata determines discoverability, use and value of its associated
data. To this end, a system for metadata management, which
\textbf{centralizes} metadata from distributed data sources, is defined
and its implementation described in the following.

This work focuses expressly on university libraries; and from a library
perspective, the overall concept corresponds to a \textbf{union catalog}
(Caplan 2003, Ch.~4) of publications and research data. Considering then
that \emph{data operations} (DataOps) is intertwined with metadata
management (Reis and Housley 2022, Ch.~6; Strengholt 2023, Ch.~9),
operationalizing research data founds upon enabling libraries to manage
distributed research data's metadata.

A starting point is the employed abstract data architecture: A
\textbf{data-lake} is a central repository for the storage of (raw)
data, typically together with at least some facility to ingest data
(Mathis 2017; Serra 2024). The rationale for considering a data-lake is
that the original raw (meta)data is still available for custom
applications. Compared to a \textbf{data warehouse}, a data-lake stores
not (just) structured data, but also semi-structured and unstructured
data.

To make a data-lake navigable and prevent deterioration into a
\emph{data swamp}, an associated \textbf{metadata catalog} indexes the
stored data by its metadata, as illustrated in the ``Functional
architecture of a data lake'' (Laurent and Madera 2020, Ch.~1). Such
metadata management in data-lakes is described in Sawadogo et al. (2019)
and references therein. Additionally, see Boukraa, Bala, and Rizzi
(2024) for a review of metadata management in data-lakes.

This work discusses a special case of the data-lake: the
\textbf{metadata-lake}\footnote{One of the first mentions of the term
  ``metadata-lake'' is in Sankar (2021).}; it's admissible data is
constrained to metadata encoded in plain-text.

Given a setting where such a metadata-lake ingests metadata from
distributed data sources, it induces a \textbf{virtual data-lake} (see
Section~4) due to the unified access, while being its own metadata
catalog, cf. Serra (2024), Ch.~6.

A metadata-lake can also be seen as a \textbf{data catalog}, as it
represents an inventory of available datasets. However, a data catalog
typically builds a database from metadata fields in linked databases
(such as column descriptions), while the metadata-lake incorporates
metadata from records in linked databases (such as bibliographic
metadata). Another related metadata system is the \textbf{data map}
(Rooney et al. 2021), which differs by focusing on tracking metadata
changes, whereas the metadata-lake provides an aggregation of current
metadata records.

The foundational data structure of the metadata-lake is the
entity-relationship model, where entity attributes are handled as a
``metadata repository supporting multiple formats without record
conversion'' (Zeng and Qin 2022, Ch.~8.5.2), while the entity relations
are handled as ``aggregation and enriched metadata records in a
repository'' (Zeng and Qin 2022, Ch.~8.5.3).

Beyond formally defining a metadata-lake, also the (open-source)
implementation of the metadata-lake
``\textbf{\texttt{DatAasee}}''\footnote{The Aasee is an artificial lake
  in Münster, Germany.} (\url{https://github.com/ulbmuenster/dataasee})
is described. \texttt{DatAasee} is designed to centralize and unify
bibliographic data as well as metadata of research data and aimed at
universities and research institutes, with explicit application at the
University and State Library of Münster.

Overall, this work is organized as follows: In Section~2 the potential
use-cases for a metadata-lake are detailed. Section~3 reviews the types
of metadata and derives a data model for metadata. Then in Section~4 an
abstract definition for a metadata-lake is constructed, while in
Section~5 the metadata-lake implementation \texttt{DatAasee} is
described. Finally, Section~6 briefly evaluates \texttt{DatAasee} in
terms of existing metadata system features and also proposes alternative
criteria.

\hypertarget{use-cases}{%
\section{Use-Cases}\label{use-cases}}

Before describing the metadata-lake, six potential use-cases and
applications for such a service are summarized. All of these were motive
for investigating the metadata-lake concept and implementing the
\texttt{DatAasee} metadata-lake.

\hypertarget{centralized-metadata}{%
\subsection{Centralized Metadata}\label{centralized-metadata}}

First and foremost, the purpose of the proposed metadata-lake is to
centralize metadata from multiple sources, to provide a single point of
access in a uniform format for university-wide (or company-wide)
metadata and thus ultimately data, cf. Anderson (2022). This means
distributed data is made available through its metadata by
data-pipelining into a central repository, which is the central theme in
data engineering (Reis and Housley 2022).

\hypertarget{cross-discipline-discoverability}{%
\subsection{Cross-Discipline
Discoverability}\label{cross-discipline-discoverability}}

A related, but mainly academic motivation is interdisciplinary data
discovery. Facilitating findability of datasets without knowledge of
discipline-specific metadata formats requires cataloging the metadata in
a common format. Aiding non-specialists, the metadata-lake partially
transforms ingested metadata to a uniform schema which is
discipline-agnostic.

\hypertarget{metadata-hub}{%
\subsection{Metadata Hub}\label{metadata-hub}}

While the previous applications focus on manual or semi-manual queries,
a further use-case of the metadata-lake is to function as a data-hub for
metadata-provisioning of other systems. In this setting all
local\footnote{w.r.t. to the organizational unit the metadata-lake
  spans.} metadata processing systems access metadata via the
metadata-lake, which entails simplifications in their implementation and
operation due to hiding heterogeneous source systems behind an
intermediate metadata-layer.

\hypertarget{fair-compliant-repositories}{%
\subsection{FAIR-Compliant
Repositories}\label{fair-compliant-repositories}}

The \textbf{FAIR} (Findable, Accessible, Interoperable, Reusable)
principles (Wilkinson 2016) formalize good scientific practices
discipline-agnostically. However, even if a particular research artifact
is FAIR, research in general can only be FAIR if it, and particularly
its metadata, is managed in a repository supporting FAIRness.

In Castro et al. (2022), a software reference architecture for general
FAIR-compliant repositories is proposed. This architecture includes a
so-called ``metadata storage layer'', which could be realized by the
metadata-lake proposed in this work\footnote{Note that in Castro et al.
  (2022) the metadata storage layer contains a metadata-lake component,
  yet this work's metadata-lake corresponds to the full ``metadata
  storage layer''.}. Accordingly, the metadata of research data objects
from all ``personal repositories'' would be gathered in a metadata-lake,
transformed and made accessible as well as associated research data
retrievable through their identifiers.

\hypertarget{dataspace}{%
\subsection{Dataspace}\label{dataspace}}

A dataspace is a data management ansatz based on a set of data sources
and relations among their contents, as well as a common interface for
accessing the participating data source's contents agnostic of their
format (Curry 2020). By unifying data access to distributed data sources
without integrating the data itself, conceptually a dataspace is a
graph-based catalog for a virtual data-lake. In this sense, the herein
proposed metadata-lake can instantiate such a dataspace.

\hypertarget{funding-compliance}{%
\subsection{Funding Compliance}\label{funding-compliance}}

Catalogs of metadata could become required by funding agencies, for
example in Germany, as indicated in a government white paper about a
planned research data law (Bundesministerium für Bildung und Forschung
2024), where the stated associated goal is improved research data
discoverability. In this scenario public research institutes have to
operate a metadata cataloging service, such as a metadata-lake, in order
to comply with funding prerequisites.

\hypertarget{metadata}{%
\section{Metadata}\label{metadata}}

As this work is at the intersection of database software and research
data or bibliographic metadata, the meaning of the term ``metadata''
needs to be defined in this context. Combining definitions from
Pomerantz (2015) and Baca (2016), metadata is set to mean here:

\textbf{Definition~1 (Metadata):}

\begin{quote}
All statements about a (tangible or digital) information object are
\textbf{metadata}.
\end{quote}

For instance, metadata of a database is, in case of a relational
database, a description of a table column. On the other hand, metadata
about research data as well as bibliographic metadata is typically
understood as a set of standardized key-value pairs describing the
underlying data or published resource. The latter is the sole
\textbf{data} of the metadata-lake and object of study in this work,
which henceforth is referred to by the term ``metadata'' unless noted
otherwise.

Following Sawadogo et al. (2019) and Ravat and Zhao (2019), metadata can
be organized into three categories: intra-object metadata, inter-object
metadata, and global metadata. Each of which is detailed next. Note,
that intra-object metadata and inter-object metadata correspond to
attributes and relations in the entity-relationship model (Zeng and Qin
2022, Ch.~4.1.1).

\hypertarget{intra-object-metadata}{%
\subsection{Intra-Object Metadata}\label{intra-object-metadata}}

Intra-object metadata refers to all literal values describing an
underlying dataset. In the classic case of bibliographic metadata it is
comprised of (among others) title, creator(s), publisher, publication
date, type, identifier, etc. Given, that a (meta-)data processing
service, such as a metadata-lake, also amounts metadata with respect to
its data - the metadata, a more fine-grained classification of
intra-object metadata types is advised. For the specific application of
the metadata-lake the following classes are employed:

\begin{itemize}
\item
  \textbf{Descriptive Metadata}, also known as business metadata, refers
  to information about the contents of a dataset, such as the
  bibliographic metadata mentioned above, and is denoted subsequently by
  \(m_{\text{descriptive}}\).
\item
  \textbf{Technical Metadata} also known as structural metadata,
  describes properties of the metadata's underlying dataset (i.e.~a file
  or archive of files) such as location, format, size, or checksum, and
  is denoted by \(m_{\text{technical}}\).
\item
  \textbf{Processual Metadata}, also known as operational metadata,
  encompasses information about the metadata record's source, internal
  identifier, or creation and modification dates, and is symbolized by
  \(m_{\text{processual}}\).
\item
  \textbf{Administrative Metadata} determines access eligibility for the
  metadata itself as well as the underlying dataset, and is denoted by
  \(m_{\text{administrative}}\).
\item
  \textbf{Social Metadata} (Yassin 2022; Strengholt 2023, Ch.~9) refers
  to metadata related to discoverability and usage, such as keywords
  from controlled and uncontrolled vocabularies or number of views, and
  is denoted by \(m_{\text{social}}\).
\item
  \textbf{Raw Metadata} refers to the originally ingested metadata from
  other data sources, justifies the data-lake aspect of the
  metadata-lake, and is denoted by \(m_{\text{raw}}\).
\end{itemize}

Altogether, intra-object metadata \(m_{\text{intra}}\) is the union of
these classes: \[
m_{\text{intra}} := \{ m_{\text{descriptive}}, m_{\text{technical}}, m_{\text{processual}}, m_{\text{administrative}}, m_{\text{social}}, m_{\text{raw}} \}.
\] Among these none is generally elevated, yet \(m_{\text{raw}}\) is
defining for this work. For instance, metadata management using
intra-object metadata without \(m_{\text{raw}}\) yields a data
warehouse, while using intra-object metadata of exclusively
\(m_{\text{raw}}\) results in an unmanaged data-lake.

\hypertarget{inter-object-metadata}{%
\subsection{Inter-Object Metadata}\label{inter-object-metadata}}

While intra-object metadata collects literal data about a dataset,
inter-object metadata describes the relation between underlying
datasets. Following Sawadogo et al. (2019), relations between datasets
are classified by:

\begin{itemize}
\item
  \textbf{Grouping} relates datasets where either one is part of another
  or both are part of a third, and denoted by \(m_{\text{grouping}}\).
\item
  \textbf{Similarity} relates datasets which are, the same work,
  manifestation, expression, or item using the terminology of FRBR
  (Functional Requirements for Bibliographic Records 2009); and is
  symbolized by \(m_{\text{similarity}}\).
\item
  \textbf{Parenthood} relates datasets where one descends from the
  other, and is denoted by \(m_{\text{parenthood}}\).
\end{itemize}

These interrelations between datasets are metadata which is represented
by relations between metadata records, and jointly inter-object metadata
is represented by \(m_{\text{inter}}\) as union of these classes: \[
m_{\text{inter}} := \{ m_{\text{grouping}}, m_{\text{similarity}}, m_{\text{parenthood}} \}.
\]

\hypertarget{global-metadata}{%
\subsection{Global Metadata}\label{global-metadata}}

Lastly, global metadata refers in this context to metadata about the
metadata records, which corresponds to the metadata understanding of
databases and potentially includes:

\begin{itemize}
\item
  \textbf{Database schema} modeling the entities by defining and
  constraining attributes and relations.
\item
  \textbf{Enumerated vocabularies} listing standardized values used
  exclusively in certain attributes.
\item
  \textbf{Context information}, such as database comments, for instance
  in SQL by \texttt{COMMENT\ ON\ DATABASE} (PostgreSQL Global
  Development Group n. d.).
\item
  \textbf{Labels} and \textbf{descriptions} which provide human-readable
  phrases for attributes, relations and content details, with use, for
  example, in a frontend application; cf. Brickley and Guha (2014).
\end{itemize}

Global metadata is rather implementation-specific and not of interest in
the subsequent section.

\hypertarget{metadata-lake}{%
\section{Metadata-Lake}\label{metadata-lake}}

This section provides a formal definition of the metadata-lake, which is
based upon its foundational data architecture - the data-lake.

\textbf{Definition~2 (Data-Lake)}

\begin{quote}
A \textbf{data-lake} is a set of datasets.
\end{quote}

Note, that the data may or may not be in some format (schema).
Furthermore, since a set consists of different elements, it needs to be
ensured that all elements are distinguishable (Maccioni and Torlone
2018), for example by a unique (but local, w.r.t. the metadata-lake)
record identifier, which cannot be an identifier of the underlying
dataset, i.e.~a DOI (Digital Object Identifier), since a dataset maybe
listed in multiple source systems.

Now, given the aim of this work, a metadata-lake cannot just be defined
as a set of metadata records; beyond holding a record identifier and the
original raw metadata, an internal metadata-lake format needs additional
fields to provide a metadata catalog and fulfill the previously
presented use-cases. Particularly, a field is needed to locate the
metadata record's underlying data if the metadata-lake is used as
metadata catalog for a virtual data-lake.

Such a schema is derived using the metadata classification from
Section~3. Overall, the metadata-lake is made up of a hierarchy of
tuples:

\textbf{Definition~3 (Intra-Object Attributes):}

\begin{quote}
A set of ordered pairs \(\{(K,V)_i\}\) is called \textbf{intra-object
attributes} if:

\begin{enumerate}
\def\labelenumi{\arabic{enumi}.}
\tightlist
\item
  \(K_i \in L_{\text{intra}}\), with \(L_{\text{intra}}\) being a finite
  set of labels,
\item
  there exists a map \(f:\{V_i\} \to L_{\text{intra}}\) such that
  \(\{K_i\} = \operatorname{image}(f)\).
\end{enumerate}
\end{quote}

Intra-object attributes conform to Definition~1, since one can
implicitly insert the word ``is'' between a key and its value, and a
statement is created of the form ``\(K_i\) is \(V_i\)''. In practical
technical terms intra-object metadata is a special case of a key-value
document, typical for document databases.

If an internal identifier is used to ensure discriminability, it needs
to be represented by an intra-object attribute, and similarly the raw
metadata needs to be encoded by such a pair. Furthermore, an external
identifier (i.e.~a data location) should be included to actually span a
virtual data-lake.

Since a common format for all metadata records is assumed, a space of
all possible valid sets of such pairs can be called
\(M_{\text{intra}}\), and the associated set of labels
\(L_{\text{intra}}\) is abstractly defined in Section~3.1.

\textbf{Definition~4 (Inter-Object Relations):}

\begin{quote}
A set of ordered triples \(\{(A,R,B)_i\}\) is called
\textbf{inter-object relations} if:

\begin{enumerate}
\def\labelenumi{\arabic{enumi}.}
\tightlist
\item
  \(R_i \in L_{\text{inter}}\), with \(L_{\text{inter}}\) being a finite
  set of labels.
\item
  \(A_i, B_i \in M_{\text{intra}}\)
\end{enumerate}
\end{quote}

This is equivalent to the serialization of a directed labeled graph,
where \(M_{\text{intra}}\) is the set of nodes, the ordered pairs
\((A_i, B_i)\) form the edges, and some map
\(g:\{(A_i, B_i)\} \to L_{\text{inter}}\) exists such that
\(\{R_i\} = \operatorname{image}(g)\).

If the labels conform to the (usual) constraint of being a predicate,
then also by ``\(A_i\) \(R_i\) \(B_i\)'' statements are given,
cf.~Semantic Triple (Wikipedia contributors 2024), and the inter-object
relations fit the \emph{Metadata} Definition~1.

Similarly, the space of all valid triples \(M_{\text{inter}}\) is
declared, and the set of labels \(L_{\text{inter}}\) is abstractly
defined in Section~3.2.

\textbf{Definition~5 (Metadata Graph):}

\begin{quote}
A graph \(G\) where the vertices are a set of intra-object attributes
\(m_{\text{intra}} \subset M_{\text{intra}}\), and the edges are a set
of inter-object relations \(m_{\text{inter}} \subset M_{\text{inter}}\),
connecting the former pairwise, is called (in this context)
\textbf{metadata graph},
\end{quote}

\[
G := (m_{\text{intra}}, m_{\text{inter}}).
\]

With respect to Sawadogo et al. (2019), a metadata-lake can then be
defined by:

\textbf{Definition~6 (Metadata-Lake):}

\begin{quote}
A \textbf{metadata-lake} is an ordered triple \((G,T,S)\), where \(G\)
is a metadata graph, \(T\) is a set of transformations, and \(S\) is a
set of sources, such that \(G := \bigcup_{i=1}^{|S|} T_{f(i)}(S_i)\) for
some suitable map \(f : \{1 \dots |S|\} \to \{1 \dots |T|\}\), with
\(|\cdot|\) denoting a set's cardinality.
\end{quote}

This definition is closely related to the data-lake model in Diamantini
et al. (2018), Sec.~3, with the distinction that the sources \(S\) are
metadata sources instead of data sources (\(D\)), and the addition of
transformations \(T\). Also, the ``metalake'' from Strengholt (2023),
Ch.~9 is a related concept.

\begin{figure}
\centering
\includegraphics[width=\textwidth,height=0.4\textheight]{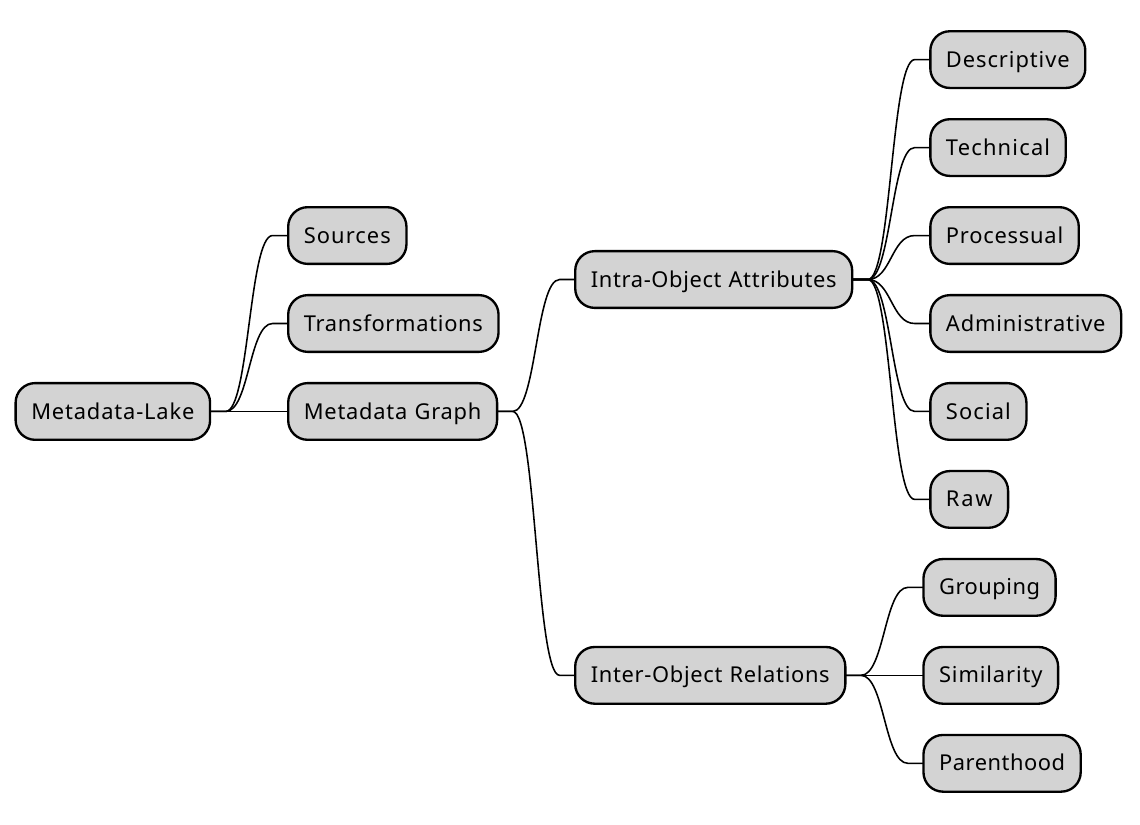}
\caption{Abstract metadata-lake.\label{hierarchy}}
\end{figure}

This abstract definition of a metadata-lake induces already all
components for a practical implementation. First, some kind of
database\footnote{The downstream access to the stored metadata can be
  realized as a semantic layer.} to hold the metadata graph \(G\),
second, a processor providing transformations \(T\) from source formats
to the metadata-lake's internal format, and third, a configuration
holding an encoding of the sources \(S\), so that the contents of \(S\)
can be transformed by \(T\) and included into \(G\).
Figure~\ref{hierarchy} illustrates the full hierarchy of the
metadata-lake.

Note, that the lake aspect of the metadata-lake is given by the raw
(meta)data \(m_{\text{raw}}\). This means, assuming a common minimal
encoding of metadata in all sources and \(M_{\text{intra}}\), and all
transformations \(T\) are the identity map, one would have a plain
data-lake.

\pagebreak

\hypertarget{dataasee}{%
\section{\texorpdfstring{\texttt{DatAasee}}{DatAasee}}\label{dataasee}}

\begin{figure}
\centering
\includegraphics{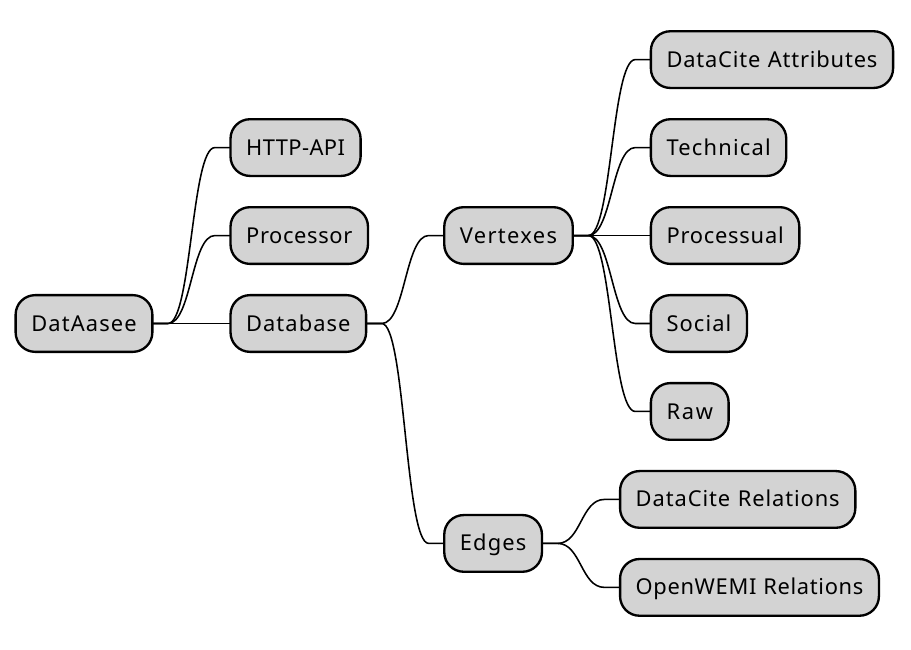}
\caption{\texttt{DatAasee} metadata-lake.\label{implementation}}
\end{figure}

Here the metadata-lake template derived in the previous section is
instantiated. This means specifying an internal metadata format within a
technical system. To this end, the following non-essential guidelines
were predetermined:

\begin{itemize}
\tightlist
\item
  A \textbf{functional software architecture} with minimal state.
\item
  Components are isolated in \textbf{containers}.
\item
  Components are implemented as \textbf{declarative} as possible.
\item
  Interaction is exclusively realized through an \textbf{HTTP-API}.
\item
  All API communication is in \textbf{JSON} encoding.
\end{itemize}

As stated earlier, the primary goal of the metadata-lake is centralizing
metadata access, and since the metadata-lake is a data-lake and at the
same time also a metadata catalog, the ``Hub and Spoke Metadata
Architecture'' from Laurent and Madera (2020), Sec.~4.8.2 is abstractly
employed. Its weakness of a customized standard meta model is mitigated
by only partially parsing ingested metadata records and thus reducing
the extent of associated computational load.

\begin{figure}
\centering
\includegraphics[width=\textwidth,height=1\textheight]{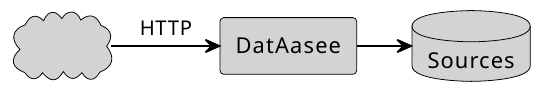}
\caption{\texttt{DatAasee} setting.\label{outward}}
\end{figure}

In a top-down approach the metadata-lake architecture is subsequently
staked out in ever more detail. For an overview of the practical data
model (see also Section~5.2.1.1) and the application of the
metadata-lake, see Figure~\ref{implementation} and Figure~\ref{outward}.

\hypertarget{data-architecture}{%
\subsection{Data Architecture}\label{data-architecture}}

The most foundational decision about building a (metadata management)
service is its data architecture. Fundamentally, the data architecture
is fixed already as data-lake, but only to the degree of Definition~2.
Implementing a data-lake requires additional practical architectural
considerations (Hai et al. 2023).

\hypertarget{data-lake-architecture}{%
\subsubsection{Data-Lake Architecture}\label{data-lake-architecture}}

A process pattern that is closely related to data-lakes is \textbf{ELT}
(Extract-Load-Transform), which summarizes a process that extracts data
from a source system, loads it as-is into the destination system, and
then transforms it on-demand. On the other hand, the corresponding
process pattern in data warehouses is \textbf{ETL}
(Extract-Transform-Load), where data is extracted from a source system,
transformed according to the destination system's schema, and loaded
into it. The metadata-lake stores raw metadata like a data-lake, as well
as transformed data like a data warehouse due to its metadata catalog
aspect. Hence, the \textbf{EtLT} (Extract-transform-Load-Transform)
process patterns from Densmore (2021) is effectively used. This means
the raw metadata is partially transformed to match the metadata-lake's
catalog schema, while retaining the original raw metadata which is then
transformed to its (external) application upon request.

Given typical data-lake architectures (Hlupić et al. 2022), the
metadata-lake corresponds to the two-layered architecture with a
transient landing zone and a second zone that divergingly not only holds
the raw (meta)data, but jointly raw and refined metadata. This directly
results from the EtLT process choice.

\hypertarget{software-architecture}{%
\subsection{Software Architecture}\label{software-architecture}}

\begin{figure}
\centering
\includegraphics[width=\textwidth,height=1\textheight]{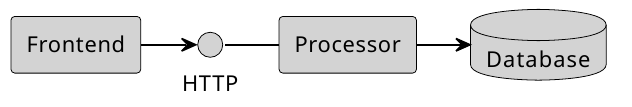}
\caption{\texttt{DatAasee} architecture.\label{inward}}
\end{figure}

\pagebreak

The foregoing data architecture is implemented following a classic
three-tier software architecture (Manuel and AlGhamdi 2003), with the
typical tiers:

\begin{itemize}
\tightlist
\item
  Data-tier: A database management system (DBMS),
\item
  Application-tier: A stateless processing system encoding the domain
  logic,
\item
  Presentation-tier: A semantic layer as interface for users or clients,
\end{itemize}

which are specified next; see also Figure~\ref{inward}.

\hypertarget{data-tier}{%
\subsubsection{Data-Tier}\label{data-tier}}

Similar to Prabhune et al. (2016), \texttt{DatAasee} is based on a
\textbf{NoSQL} database system, which stores all types of metadata
summarized in Section~3. Here the inter-object relations suggest support
of a graph model (Ravat and Zhao 2019), while the intra-object
attributes imply a document (or relational) model. A property-graph
database matches both needs by using documents as nodes and edges. Thus,
the proposed metadata model translates as follows: The intra-object
attributes correspond to the nodes (vertices), and the inter-object
relations are realized by edges, cf.~entity-relationship model.

Practically, the DBMS \textbf{ArcadeDB} (Garulli et al. n. d.) is
employed in \texttt{DatAasee}. ArcadeDB is a multi-model \textbf{NoSQL}
database featuring a hierarchy of models: An overarching graph model
with a document model for nodes and edges, and a key-value model for
document contents, resulting in a property-graph. A specialty of
ArcadeDB is its polyglotism: While the native query language is a
dialect of \textbf{SQL}, (Open-)\textbf{Cypher}, \textbf{MQL} (Mongo),
\textbf{GraphQL}, and (a subset of) \textbf{Redis} commands are also
supported.

\hypertarget{data-model}{%
\paragraph{Data Model}\label{data-model}}

The data model of the \texttt{DatAasee} data-tier follows the
metadata-lake schema proposed in Section~3 and Section~4: Intra-object
metadata is stored in a vertex, and inter-object metadata is encoded by
edges. The vertices consist of descriptive, technical, processual,
social, and raw metadata, where the descriptive metadata key-value pairs
are derived from the DataCite 4.5 metadata kernel (DataCite Metadata
Working Group 2024). Furthermore, administrative metadata is omitted,
since all metadata records are unrestrictedly available to all users for
reading (license and rights fields are kept as descriptive metadata).

In terms of a relational database, the \texttt{metadata} vertex schema
(intra-object attributes) corresponds to a wide table with a star schema
for enumerable dimensions. Edge types are a subset of the DataCite item
relations together with a subset of the OpenWEMI (Dublin Core Metadata
Initiative 2024) entity relations. For details about the database
schema, see the Appendix.

In the metadata-lake, the two mandatory intra-object attributes are the
record identifier and raw metadata. While the former is computed as a
\textbf{SHA256} hash from a subset of descriptive and processual
metadata and then \textbf{base64url} encoded, the latter is stored as a
plain string, since only textual metadata was assumed.

\pagebreak

\hypertarget{alternative-implementations}{%
\paragraph{Alternative
Implementations}\label{alternative-implementations}}

The property-graph-based implementation chosen for \texttt{DatAasee} is
not imperative. Likewise, a relational or knowledge-graph-based
data-tier could have been used. In case of a relational database, the
inter-object metadata would be emulated by a single or multiple junction
tables, while in a knowledge graph, the intra-object metadata can be
constructed via statements with literal objects. Thus, if implementation
constraints or additional requirements demand it, alternative database
types are compatible. However, in the author's opinion a property graph
is the ideal choice for the proposed metadata-lake.

\hypertarget{application-tier}{%
\subsubsection{Application-Tier}\label{application-tier}}

Basically, the domain logic consists of two components, the import of
metadata on the one hand, and the provisioning of it on the other. The
import from source systems includes the request-driven transport,
parsing, (partial) transformation, data-cleaning, metadata quality
evaluation and storing in the data-tier. Provisioning happens by
listening for presentation-tier requests, translating those to data-tier
queries, and re-translating the results to a domain response, akin to a
semantic layer. Requests for a single record by its identifier or search
by filter, full-text or custom query in a compatible query-language are
supported.

\hypertarget{data-pipelines}{%
\paragraph{Data Pipelines}\label{data-pipelines}}

The source and transformation components of the metadata-lake, with
itself as implicit destination, form data pipelines. A data pipeline
needs to be specialized by a configuration to be able to admit updates
or additions during operation. Technically, four (metadata) items are
minimally needed to specify a data source and transformation for
\texttt{DatAasee}:

\begin{itemize}
\tightlist
\item
  \textbf{Location}: is a string that translates to a resolvable
  address, i.e.: \textbf{URL}.
\item
  \textbf{Protocol}: specifies sequence and content of messages, i.e.:
  \textbf{OAI-PMH}.
\item
  \textbf{Encoding}: fixes the manner of metadata payload formulation,
  i.e.: \textbf{XML}.
\item
  \textbf{Format}: then describes how the content is arranged, i.e.:
  \textbf{DataCite}.
\end{itemize}

Location and protocol are required to \emph{extract} the metadata from
its source, while encoding and format are essential to (partially)
\emph{transform} the metadata into the native metadata-lake schema.
Potentially, also credentials (i.e.: username/password,
access-/secret-key) maybe necessary to gain access to a source.
Non-technically, \texttt{DatAasee} requires a data-steward to be named
for every ingested source, as well as the rights (license) under which
the metadata of the source are ingested. Beyond these, an options field
is available to pass protocol specific configuration, i.e.~selective
harvesting for OAI-PMH.

Currently, \texttt{DatAasee} supports the \textbf{OAI-PMH}, \textbf{S3},
and \textbf{GET} protocols, as well as \textbf{DataCite},
\textbf{DublinCore}, \textbf{LIDO}, \textbf{MARC}, and \textbf{MODS}
formats\footnote{See
  \href{https://www.openarchives.org/pmh}{openarchives.org/pmh},
  \href{https://docs.aws.amazon.com/s3}{docs.aws.amazon.com/s3},
  \href{https://httpwg.org/specs/rfc9110\#GET}{httpwg.org/specs/rfc9110},
  \href{https://www.dublincore.org}{dublincore.org},
  \href{https://lido-schema.org}{lido-schema.org},
  \href{https://www.loc.gov/standards/marcxml}{loc.gov/standards/marcxml},
  and \href{https://www.loc.gov/standards/mods}{loc.gov/standards/mods}}.
Encoding-wise \textbf{XML} is implicitly assumed, but is principally
extensible to \textbf{JSON}, \textbf{CSV} or others. Additionally, the
contents of another \texttt{DatAasee} instance can be ingested.

For exporting metadata, \texttt{DatAasee} provides its native format
which is derived from DataCite, as well as actual DataCite, and BibJSON
(MacGillivray and Pitman n. d.) - a JSON representation of
{\rm B\kern-.05em{\sc i\kern-.025em b}\kern-.08emT\kern-.1667em\lower.7ex\hbox{E}\kern-.125emX}.

\hypertarget{presentation-tier}{%
\subsubsection{Presentation-Tier}\label{presentation-tier}}

As an API-based web-service, cf.~Data Commons (Grossman et al. 2016),
the primary front-end of the \texttt{DatAasee} metadata-lake is a
REST-like (REpresentational State Transfer) API (Application Programming
Interface) with CQRS (Command-Query-Responsibility-Segregation) aspects
over HTTP. While the ingest encoding is XML, the HTTP-API only accepts
and responds exclusively in \textbf{JSON}. As a secondary effect,
\texttt{DatAasee} can also serve as translator from legacy XML sources
to JSON. The response content is formatted following the JSON-API (Katz
et al. 2022) standard. And to establish self-documenting messages, each
response contains a link to its content's associated \textbf{JSON
Schema}, such that the wrapping JSON-API and its payload data are
understandable. Furthermore, this API is documented via an OpenAPI
(Miller et al. 2020) document which is provided by the API itself. Since
all messages to and from the \texttt{DatAasee} API server are in JSON
encoding, for all requests and responses JSON Schemas are provided
through the API itself.

\hypertarget{web-frontend}{%
\paragraph{Web-Frontend}\label{web-frontend}}

In addition, to the primary HTTP-API interface, a secondary prototype
graphical user interface for the metadata-lake in the form a web
application is available, which in turn exclusively uses the HTTP-API.
While optional and fully decoupled from the metadata-lake proper, and
beyond being a demonstrator facilitating exploration and review, it
serves the purposes of:

\begin{itemize}
\tightlist
\item
  template implementation of an exemplary graphical user interface;
\item
  living documentation, for a hands-on interaction with the
  metadata-lake;
\item
  more comfortable manual testing, and abstracting lower-level
  formatting.
\end{itemize}

\hypertarget{preliminary-evaluation}{%
\section{Preliminary Evaluation}\label{preliminary-evaluation}}

There are various evaluation criteria for metadata systems of
data-lakes, such as the functional features: Semantic Enrichment, Data
Indexing, Link Generation, Data Polymorphism, Data Versioning, and Usage
Tracking from Sawadogo et al. (2019), the generic features: Metadata
Flexibility, Granular Levels, and Multiple Zones from Eichler et al.
(2020), or the semantic features: Semantic Labels, Semantic
Relationships, Central vs External Graph, Metadata Interoperability,
Initial Automatic Labeling, Technical Abstraction, and Open-Source from
Hoseini, Theissen-Lipp, and Quix (2023). Notably, these features are
primarily concerned with the internal organization of the metadata
management. In Table~1, a brief evaluation of \texttt{DatAasee} in terms
of the aforementioned features is summarized; for a description of these
features see the respective references.

\pagebreak

\begin{longtable}[]{@{}
  >{\raggedleft\arraybackslash}p{(\columnwidth - 4\tabcolsep) * \real{0.4912}}
  >{\centering\arraybackslash}p{(\columnwidth - 4\tabcolsep) * \real{0.2807}}
  >{\raggedright\arraybackslash}p{(\columnwidth - 4\tabcolsep) * \real{0.2281}}@{}}
\caption{Overview of \texttt{DatAasee} Features}\tabularnewline
\toprule\noalign{}
\begin{minipage}[b]{\linewidth}\raggedleft
\textbf{Feature}
\end{minipage} & \begin{minipage}[b]{\linewidth}\centering
\textbf{\texttt{DatAasee}}
\end{minipage} & \begin{minipage}[b]{\linewidth}\raggedright
\textbf{Comment}
\end{minipage} \\
\midrule\noalign{}
\endfirsthead
\toprule\noalign{}
\begin{minipage}[b]{\linewidth}\raggedleft
\textbf{Feature}
\end{minipage} & \begin{minipage}[b]{\linewidth}\centering
\textbf{\texttt{DatAasee}}
\end{minipage} & \begin{minipage}[b]{\linewidth}\raggedright
\textbf{Comment}
\end{minipage} \\
\midrule\noalign{}
\endhead
\bottomrule\noalign{}
\endlastfoot
Semantic Enrichment & Yes & Schema includes categorization \\
Data Indexing & Yes & Data-tier runs (full-text) indices \\
Link Generation & Yes & Based on relations and identifiers \\
Data Polymorphism & Yes & Raw and refined metadata stored \\
Data Versioning & No & Only if tracked via relations \\
Usage Tracking & No & Not yet implemented \\
Metadata Flexibility & Yes & Uses document vertices and edges \\
Granular Levels & No & Only via grouping relations \\
Multiple Zones & Yes & Raw and refined zone \\
Semantic Labels & No & Only in terms of DataCite relations \\
Semantic Relationships & Yes & Via DataCite relations \\
Central vs External Graph & No & Federation is not yet implemented \\
Metadata Interoperability & Yes & Exportable as JSON \\
Initial Automatic Labeling & Yes & Final step of ingest process \\
Technical Abstraction & No & Advanced use needs query languages \\
Compatibility & No & Only with DataCite and OpenWEMI \\
Open-Source & Yes & MIT-licensed \\
\end{longtable}

\hypertarget{fair-as-alternative-features}{%
\subsection{FAIR as Alternative
Features}\label{fair-as-alternative-features}}

An alternative set of features for the evaluation of metadata layers or
metadata systems can be derived from the FAIR principles\footnote{See
  also: \url{https://www.go-fair.org/fair-principles}}. Those are not
meant to be added, but rather as an alternative, since their coverage of
underlying qualities overlaps with the features above. In comparison,
the following criteria are concerned with the outward functionality
instead of internal structure:

\begin{itemize}
\item
  \textbf{Findability} is given by the exposed search modes, such as
  \emph{full-text} (F1), \emph{filters} or \emph{facets} (F2), or via
  \emph{query languages} (F3).
\item
  \textbf{Accessibility} is provided by an API, which should follow a
  \emph{standard} (A1), be \emph{self-explaining} (A2), and
  \emph{documented} (A3).
\item
  \textbf{Interoperability} refers to flexibility in \emph{protocols}
  (I1), \emph{encodings} (I2) and \emph{formats} (I3) from which can be
  imported and to which can be exported.
\item
  \textbf{Reusability} is enabled by open source code \emph{licensing}
  (R1), multi-layered \emph{documentation} (R2) and hardware/software
  \emph{compatibility} (R3).
\end{itemize}

All of the above are checked by \texttt{DatAasee}.

\hypertarget{summary}{%
\section{Summary}\label{summary}}

A foundational attribute for research data, research software, or
publications to be useful is first and foremost their availability (Fehr
et al. 2016). So, a data-lake - or metadata-lake - with its priority of
\emph{load} before \emph{transform} (in the context of ELT) qualifies
for the task of research metadata management. The metadata cataloging
aspect of the metadata-lake then makes it a suitable metadata layer for
a virtual data-lake, dataspace, or FAIR-compliant repository.

\texttt{DatAasee} is an open-source implementation of the proposed
metadata-lake architecture, and aims to become the metadata-layer in
academic institutions such as universities and research institutes.
Particularly, to automate metadata handling and improve research data
discoverability in university libraries, academic libraries, scientific
libraries or research libraries.

\section*{Code Availability Statement}

The source code for \texttt{DatAasee} (version 0.9) is available under
MIT-license at:

\begin{quote}
\href{https://doi.org/10.5281/zenodo.20067751}{\texttt{https://doi.org/10.5281/zenodo.20067751}}
\end{quote}

\section*{Acknowledgement}

The author thanks \href{https://orcid.org/0000-0003-1844-9722}{Holger
Przibytzin}, \href{https://orcid.org/0000-0001-9675-1893}{Marc
Schutzeichel}, and \href{https://orcid.org/0009-0004-7264-2843}{Jan-Erik
Stange} for providing valuable feedback on this manuscript.

\newpage

\section*{Appendix}

Below is the \texttt{DatAasee} data-tier database schema, encoded in
\textbf{yasql}\footnote{See:
  \href{https://github.com/aryelgois/yasql}{github.com/aryelgois/yasql}},
visualized:

\includegraphics[width=\textwidth,height=0.89\textheight]{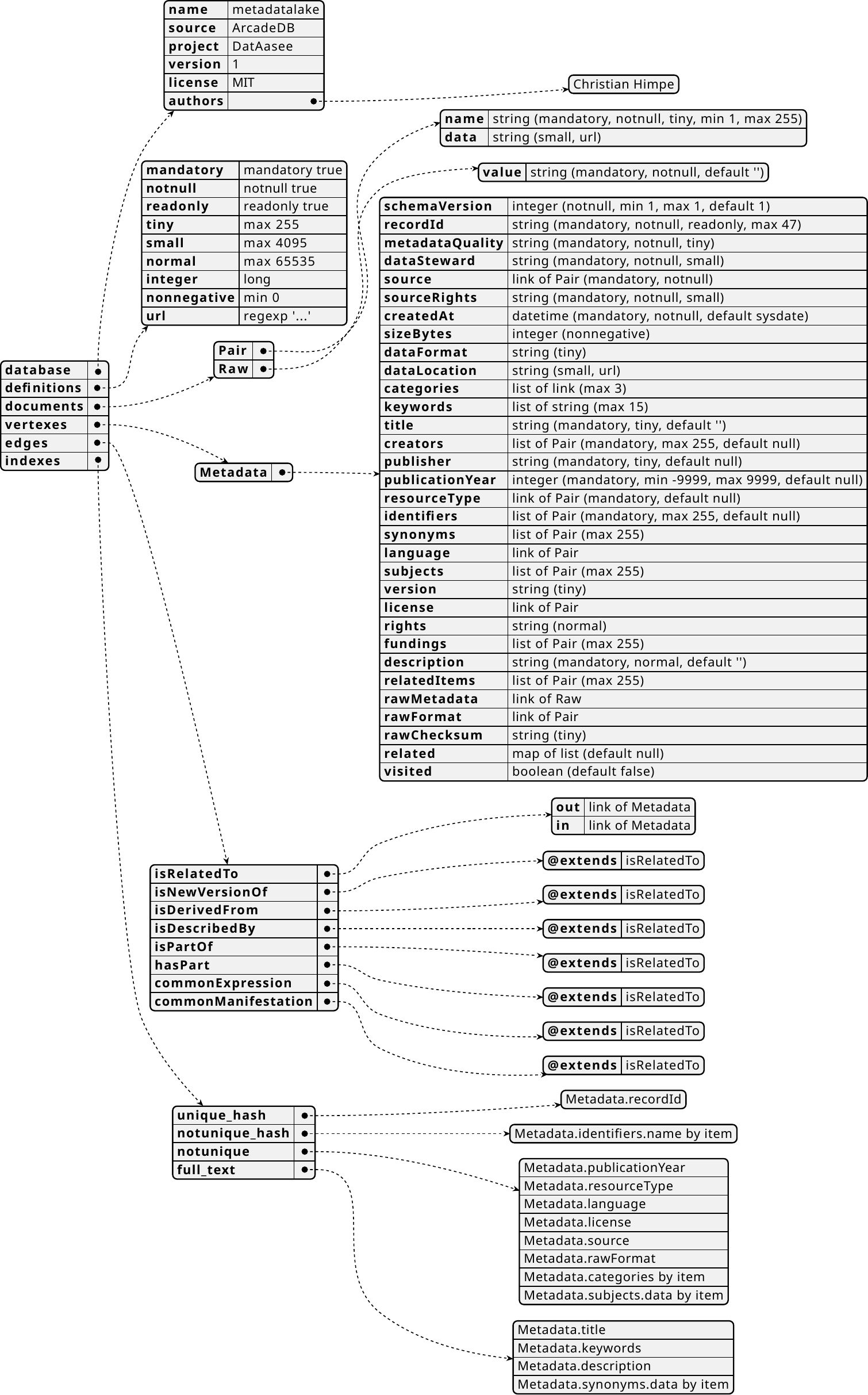}

\newpage

\section*{Bibliography}

\hypertarget{refs}{}
\begin{CSLReferences}{1}{0}
\leavevmode\vadjust pre{\hypertarget{ref-And22}{}}%
Anderson, C. B. 2022. {``\textbf{An Introduction to Data Lakes for
Academic Librarians}.''} \emph{Information Services \& Use} 42:
397--407. \url{https://doi.org/10.3233/ISU-220176}.

\leavevmode\vadjust pre{\hypertarget{ref-Bac16}{}}%
Baca, M., ed. 2016. {``\textbf{Introduction to Metadata}.''}
\url{http://www.getty.edu/publications/intrometadata}.

\leavevmode\vadjust pre{\hypertarget{ref-Bou24}{}}%
Boukraa, D., M. Bala, and S. Rizzi. 2024. {``\textbf{Metadata Management
in Data Lake Environments: A Survey}.''} \emph{Journal of Library
Metadata}, 1--60. \url{https://doi.org/10.1080/19386389.2024.2359310}.

\leavevmode\vadjust pre{\hypertarget{ref-Bri14}{}}%
Brickley, D., and R. V. Guha, eds. 2014. {``RDF Schema 1.1.''}
\url{https://www.w3.org/TR/rdf-schema/}.

\leavevmode\vadjust pre{\hypertarget{ref-Bun24}{}}%
Bundesministerium für Bildung und Forschung. 2024. {``\textbf{Eckpunkte
{BMBF} Forschungsdatengesetz}.''} {[}German{]}
\url{https://www.bmbf.de/SharedDocs/Downloads/DE/gesetze/forschungsdatengesetz/sonstige/Eckpunktepapier.pdf?__blob=publicationFile\&v=3}.

\leavevmode\vadjust pre{\hypertarget{ref-Cap03}{}}%
Caplan, P. 2003. \emph{\textbf{Metadata Fundamentals for All
Librarians}}. ALA.
\url{https://alastore.ala.org/content/metadata-fundamentals-all-librarians}.

\leavevmode\vadjust pre{\hypertarget{ref-Cas22}{}}%
Castro, J. P. C., L. M. F. Romero, A. C. Carniel, and C. D. Aguiar.
2022. {``\textbf{{FAIR} Principles and Big Data: A Software Reference
Architecture for Open Science}.''} In \emph{24th International
Conference on Enterprise Information Systems - Volume 1: {ICEIS}},
27--38. \url{https://doi.org/10.5220/0011045500003179}.

\leavevmode\vadjust pre{\hypertarget{ref-Cur20}{}}%
Curry, E. 2020. {``\textbf{Dataspaces: Fundamentals, Principles, and
Techniques}.''} In \emph{Real-Time Linked Dataspaces}, 45--62.
\url{https://doi.org/10.1007/978-3-030-29665-0_3}.

\leavevmode\vadjust pre{\hypertarget{ref-Dat24}{}}%
DataCite Metadata Working Group. 2024. {``\textbf{DataCite Metadata
Schema for the Publication and Citation of Research Data and Other
Research Outputs (Version 4.5)}.''}
\url{https://doi.org/10.14454/znvd-6q68}.

\leavevmode\vadjust pre{\hypertarget{ref-Den21}{}}%
Densmore, J. 2021. \emph{\textbf{Data Pipelines Pocket Reference}}.
O'Reilly Media.
\url{https://www.oreilly.com/library/view/data-pipelines-pocket/9781492087823/}.

\leavevmode\vadjust pre{\hypertarget{ref-Dia18}{}}%
Diamantini, C., P. Lo Giudice, L. Musarella, D. Potena, E. Storti, and
D. Ursino. 2018. {``\textbf{A New Metadata Model to Uniformly Handle
Heterogeneous Data Lake Sources}.''} In \emph{{ABDIS} 2018: New Trends
in Databases and Information Systems}, 165--77.
\url{https://doi.org/10.1007/978-3-030-00063-9_17}.

\leavevmode\vadjust pre{\hypertarget{ref-Ope24}{}}%
Dublin Core Metadata Initiative, OpenWEMI Working Group. 2024.
{``\textbf{OpenWEMI}.''}
\url{https://www.dublincore.org/specifications/openwemi/specification/}.

\leavevmode\vadjust pre{\hypertarget{ref-Eic20}{}}%
Eichler, R., C. Giebler, C. Gröger, H. Schwarz, and B. Mitschang. 2020.
{``\textbf{{HANDLE} - a Generic Metadata Model for Data Lakes}.''} In
\emph{Big Data Analytics and Knowledge Discovery. {DaWaK 2020}}, 73--88.
\url{https://doi.org/10.1007/978-3-030-59065-9_7}.

\leavevmode\vadjust pre{\hypertarget{ref-Feh16}{}}%
Fehr, J., J. Heiland, C. Himpe, and J. Saak. 2016. {``\textbf{Best
Practices for Replicability, Reproducibility and Reusability of
Computer-Based Experiments Exemplified by Model Reduction Software}.''}
\emph{AIMS Mathematics} 1: 261--81.
\url{https://doi.org/10.3934/Math.2016.3.261}.

\leavevmode\vadjust pre{\hypertarget{ref-Ifl09}{}}%
Functional Requirements for Bibliographic Records, IFLA Study Group on
the. 2009. {``\textbf{Functional Requirements for Bibliographic Records:
Final Report}.''}
\url{https://repository.ifla.org/handle/20.500.14598/811}.

\leavevmode\vadjust pre{\hypertarget{ref-Arc}{}}%
Garulli, L. et al. n. d. {``\textbf{ArcadeDB}.''}
\url{https://github.com/ArcadeData/arcadedb}.

\leavevmode\vadjust pre{\hypertarget{ref-Gro16}{}}%
Grossman, R. L., A. Heath, M. Murphy, M. Patterson, and W. Wells. 2016.
{``\textbf{A Case for Data Commons: Toward Data Science as a
Service}.''} \emph{Computing in Science \& Engineering} 18 (5): 10--20.
\url{https://doi.org/10.1109/MCSE.2016.92}.

\leavevmode\vadjust pre{\hypertarget{ref-Hai23}{}}%
Hai, R., C. Koutras, C. Quix, and M. Jarke. 2023. {``\textbf{Data Lakes:
A Survey of Functions and Systems}.''} \emph{IEEE Transactions on
Knowledge and Data Engineering} 35 (12): 12571--90.
\url{https://doi.org/10.1109/TKDE.2023.3270101}.

\leavevmode\vadjust pre{\hypertarget{ref-Hlu22}{}}%
Hlupić, T., D. Oreščanin, D. Ružak, and M. Baranović. 2022.
{``\textbf{An Overview of Current Data Lake Architecture Models}.''} In
\emph{45th Jubilee International Convention on Information,
Communication and Electronic Technology ({MIPRO})}, 1082--87.
\url{https://doi.org/10.23919/MIPRO55190.2022.9803717}.

\leavevmode\vadjust pre{\hypertarget{ref-Hos23}{}}%
Hoseini, S., J. Theissen-Lipp, and C. Quix. 2023. {``\textbf{Semantic
Data Management in Data Lakes}.''} \emph{arXiv} cs.DB: 2310.15373.
\url{https://doi.org/10.48550/arXiv.2310.15373}.

\leavevmode\vadjust pre{\hypertarget{ref-Kat22}{}}%
Katz, Y., D. Gebhardt, G. Sullice, and J. Hanschke. 2022.
{``\textbf{JSON API Specification (V1.1)}.''}
\url{https://jsonapi.org/format/1.1/}.

\leavevmode\vadjust pre{\hypertarget{ref-Lau20}{}}%
Laurent, D., and C. Madera. 2020. \emph{\textbf{Data Lakes}}. John Wiley
\& Sons, Incorporated. \url{https://doi.org/10.1002/9781119720430}.

\leavevmode\vadjust pre{\hypertarget{ref-Mac18}{}}%
Maccioni, A., and R. Torlone. 2018. {``\textbf{{KAYAK}: A Framework for
Just-in-Time Data Preparation in a Data Lake}.''} In \emph{{CAiSE} 2018:
Advanced Information Systems Engineering}, 474--89.
\url{https://doi.org/10.1007/978-3-319-91563-0_29}.

\leavevmode\vadjust pre{\hypertarget{ref-Bib}{}}%
MacGillivray, M., and J. Pitman. n. d. {``\textbf{BibJSON}.''} Open
Knowledge Labs \url{https://okfnlabs.org/projects/bibjson/}.

\leavevmode\vadjust pre{\hypertarget{ref-Man03}{}}%
Manuel, P. D., and J. AlGhamdi. 2003. {``\textbf{A Data-Centric Design
for n-Tier Architecture}.''} \emph{Information Systems} 150: 195--206.
\url{https://doi.org/10.1016/S0020-0255(02)00377-8}.

\leavevmode\vadjust pre{\hypertarget{ref-Mat17}{}}%
Mathis, C. 2017. {``\textbf{Data Lakes}.''} \emph{Datenbank Spektrum}
17: 289--93. \url{https://doi.org/10.1007/s13222-017-0272-7}.

\leavevmode\vadjust pre{\hypertarget{ref-Mil20}{}}%
Miller, D., J. Whitlock, M. Gardiner, M. Ralphson, R. Ratovsky, and U.
Sarid, eds. 2020. {``\textbf{OpenAPI Specification (V3.0.3)}.''}
\url{https://spec.openapis.org/oas/v3.0.3}.

\leavevmode\vadjust pre{\hypertarget{ref-Pom15}{}}%
Pomerantz, J. 2015. \emph{\textbf{Metadata}}. MIT Press.
\url{https://mitpress.mit.edu/9780262528511/metadata/}.

\leavevmode\vadjust pre{\hypertarget{ref-Pos}{}}%
PostgreSQL Global Development Group. n. d. {``\textbf{PostgreSQL
Documentation}.''}
\url{https://www.postgresql.org/docs/current/sql-comment.html}.

\leavevmode\vadjust pre{\hypertarget{ref-Pra16}{}}%
Prabhune, A., H. Ansari, R. Keshav A. Stotzka, M. Gertz, and J. Hesser.
2016. {``\textbf{{MetaStore}: A Metadata Framework for Scientific Data
Repositories}.''} In \emph{2016 IEEE International Conference on Big
Data}, 3026--35. \url{https://doi.org/10.1109/BigData.2016.7840956}.

\leavevmode\vadjust pre{\hypertarget{ref-Rav19}{}}%
Ravat, F., and Y. Zhao. 2019. {``\textbf{Metadata Management for Data
Lakes}.''} In \emph{{ABDIS} 2019: New Trends in Databases and
Information Systems}, 37--44.
\url{https://doi.org/10.1007/978-3-030-30278-8_5}.

\leavevmode\vadjust pre{\hypertarget{ref-Rei22}{}}%
Reis, J., and M. Housley. 2022. \emph{\textbf{Fundamentals of Data
Engineering}}. O'Reilly Media.
\url{https://www.oreilly.com/library/view/fundamentals-of-data/9781098108298/}.

\leavevmode\vadjust pre{\hypertarget{ref-Roo21}{}}%
Rooney, S., L. Garcés-Erice, D. Bauer, and P. Urbanetz. 2021.
{``\textbf{Pathfinder: Building the Enterprise Data Map}.''} In
\emph{2021 {IEEE} International Conference on Big Data}, 1909--19.
\url{https://doi.org/10.1109/BigData52589.2021.9671608}.

\leavevmode\vadjust pre{\hypertarget{ref-San21}{}}%
Sankar, P. 2021. {``\textbf{The Rise of the Metadata Lake}.''}
\url{https://towardsdatascience.com/the-rise-of-the-metadata-lake-1e95127594de}.

\leavevmode\vadjust pre{\hypertarget{ref-Saw19}{}}%
Sawadogo, P. N., E. Scholly, C. Favre, E. Ferey, S. Loudcher, and J.
Darmont. 2019. {``\textbf{Metadata Systems for Data Lakes: Models and
Features}.''} In \emph{{ADBIS} 2019: New Trends in Databases and
Information Systems}, 440--51.
\url{https://doi.org/10.1007/978-3-030-30278-8_43}.

\leavevmode\vadjust pre{\hypertarget{ref-Ser24}{}}%
Serra, J. 2024. \emph{\textbf{Deciphering Data Architectures}}. O'Reilly
Media.
\url{https://www.oreilly.com/library/view/deciphering-data-architectures/9781098150754/}.

\leavevmode\vadjust pre{\hypertarget{ref-Str23}{}}%
Strengholt, P. 2023. \emph{\textbf{Data Management at Scale}}. O'Reilly
Media.
\url{https://www.oreilly.com/library/view/data-management-at/9781098138851/}.

\leavevmode\vadjust pre{\hypertarget{ref-Wik24}{}}%
Wikipedia contributors. 2024. {``\textbf{Semantic Triple ---
{Wikipedia}{,} the Free Encyclopedia}.''}
\url{https://en.wikipedia.org/w/index.php?title=Semantic_triple\&oldid=1317867149}.

\leavevmode\vadjust pre{\hypertarget{ref-Wil16}{}}%
Wilkinson, M. D. 2016. {``\textbf{The {FAIR Guiding Principles} for
Scientific Data Management and Stewardship}.''} \emph{Scientific Data}
3: 160018. \url{https://doi.org/10.1038/sdata.2016.18}.

\leavevmode\vadjust pre{\hypertarget{ref-Yas22}{}}%
Yassin, M. 2022. {``\textbf{The Role of Metadata and Metadata Lake for a
Successful Data Architecture}.''}
\url{https://web.archive.org/web/20241116054346/https://hyperight.com/the-role-of-metadata-and-metadata-lake-for-successful-data-architecture/}.

\leavevmode\vadjust pre{\hypertarget{ref-Zen22}{}}%
Zeng, Marcia Lei, and Jian Qin. 2022. \emph{\textbf{Metadata}}. Facet
Publishing.
\url{https://www.facetpublishing.co.uk/page/detail/metadata/?k=9781783305889}.

\end{CSLReferences}

\end{document}